# Concept of Multiply Connected Superconducting Tapes

George A. Levin and Paul N. Barnes

*Abstract*— The possibility of a substantial reduction of weight and size of electrical generators is the main incentive behind the effort to develop superconducting armature windings based on $Y_1Ba_2Cu_3O_{6+x}$ (YBCO) coated conductors in the form of wide tapes with large aspect ratio. The main obstacle to the application of coated superconductors in stator windings is the large losses incurred due to the ac magnetic field produced by the rotor's dc coils of the field windings. In the range of frequencies typical for aircraft generators, the hysteresis losses in wide tapes are unacceptably high. They can be reduced by dividing the YBCO layer into multiple filaments separated by non-superconducting barriers. However, the lack of current sharing between the filaments makes the conductor vulnerable to the localized defects, so that a single blockage can impede the flow of transport current through the whole length of a given filament. We present estimates of reliability as well as the magnetization losses in multiply connected superconductors. In this type of superconducting tape, a sparse network of superconducting bridges, which allows for current sharing, connects the filaments. The trade-off between the different types of losses and the connectivity requirement imposes restrictions on the number of filaments and properties of the network of bridges.

## I. Introduction

REPLACEMENT of copper wires with superconductors promises the possibility of large reductions in the size and weight of motors and generators. The maximum benefit of such reduction can be achieved in a fully superconducting machine where both the stator and rotor windings are superconducting. The main obstacle to the implementation of such designs is the very large ac loss that would occur in the superconducting stator armature winding because it is subjected to the time-varying magnetic field [1]. A significant portion of these losses is the hysteresis component.

This problem is especially acute in the second generation high temperature superconducting (HTS) wire, $Y_1Ba_2Cu_3O_{6+x}$ (YBCO) coated superconductor, that is produced in the form of wide, thin tapes [2]. Carr and Oberly have proposed a way to reduce the hysteresis loss in such tapes [3,4]. They argued that the amount of energy converted into heat could be reduced by dividing the tape into a large number of parallel stripes separated by non-superconducting resistive barriers. An experiment on small samples of YBCO films deposited on $LaAlO_3$ substrate has confirmed the validity of that suggestion [5]. Recently, the reduction of the magnetization losses in multifilament coated superconductors by about 90% in comparison with the control (uniform) sample was reported in [6,7].

This straightforward solution of the hysteresis loss problem is not without serious drawbacks. One of them is the lack of connectivity, i.e. the ability of superconducting stripes to share supercurrent should one of them be blocked, temporarily or permanently, by a localized defect or a "hot spot"[8]. In a multifilament (striated) tape without current sharing between the stripes, these defects lead to a cumulative degradation of the current-carrying capacity of the long-length conductors. In the non-striated wide tape such defects do not have a cumulative effect on the ability of the long-length conductor to carry transport current because the supercurrent can circumvent the damaged areas. The normal regime of operation for a conductor in a generator or motor is sub-critical (the *transport* current is well below critical). In the sub-critical regime the resistive connections between the stripes cannot facilitate the sharing of the transport current, because there is no potential difference between the stripes (the superconducting areas are equipotential). Therefore, the only way to achieve a degree of connectivity in the striated superconductor operating in the subcritical regime is to provide a system of discrete or continuous superconducting bridges between the stripes. The bridges, however, will increase the hysteresis loss in comparison with the fully striated tapes. Thus, a pattern of the coated superconducting tapes for ac applications has to be devised as a compromise between competing requirements – connectivity on one hand and the reduction of the net loss on the other. Below we discuss the limitations on the parameters of such multiply connected superconducting tapes.

## II. Reduction of Losses by Striation

The power loss per unit length in a flat rectangular superconducting sample of width W and negligible thickness, exposed to a time-varying magnetic field is given by [9]

$$Q_h \approx J_c W^2 B f \; ; \quad B >> B_c \qquad (1)$$

Manuscript received October 3, 2004. G.A.L. is supported by the U.S. National Research Council Senior Research Associateship Award.
G. A. Levin is with the National Research Council, 1950 Fifth Street, Bldg. 450, Wright-Patterson AFB, OH 45433. Phone: 937-255-4780; fax: 937-656-4095; e-mail: george.levin@ wpafb.af.mil.

P. N. Barnes is with the Air Force Research Laboratory, Propulsion Directorate, 1950 Fifth Street, Bldg. 450, Wright-Patterson AFB, OH 45433 USA. e-mail: paul.barnes@wpafb.af.mil.

Here $J_c$ is the critical current per unit width, $B_c = \mu_0 J_c \ln(4)/\pi$, $\mu_0 = 4\pi \cdot 10^{-7}$ H/m is the magnetic permeability of vacuum, $B$ is the amplitude of the magnetic flux density and $f$ is the frequency. Equation (1) defines the hysteresis loss, which can be reduced by dividing the superconducting layer into stripes [3] – [5]. In the multifilament sample divided into N stripes the hysteresis loss is reduced in proportion to the width of an individual stripe $W_n \approx W/N$:

$$Q_h^{st} \approx I_c W_n B f , \qquad (2)$$

where $I_c$ is the total critical current. Thus, the hysteresis loss in striated tape can be reduced by a factor of N. However, the alternating magnetic field which penetrates through the slits between the superconducting stripes induces electric field perpendicular to the stripes: $|E_\perp| \cong BfL$, where L is the length of the sample or, in a long twisted conductor, half of the twist pitch. The power per unit length of the conductor dissipated as the result of the current flowing through the normal metal connecting the stripes[3] is:

$$Q_n^{st} \propto \rho^{-1}|E_\perp|^2 d_n W = 2\frac{(BfL)^2}{\rho} d_n W . \qquad (3)$$

Here $d_n$ is the thickness of the normal metal layer, and $\rho$ is effective resistivity. This component of loss is called a coupling loss. As shown in [6], [7] the combined loss in a multifilament sample

$$Q = Q_h^{st} + Q_n^{st} \qquad (4)$$

can be substantially reduced in comparison with the non-striated sample at the sweep rate $Bf$ of the order of several Tesla/s.

### III. END-TO-END TRANSPORT CURRENT IN MULTIFILAMENT CONDUCTORS

The improvement in loss reduction has been demonstrated on relatively short samples, 10 cm and less. An outstanding question is whether the multifilament coated conductor can be manufactured in sufficient length at a reasonable cost, comparable to that of the conventional uniform tapes. The origin of this concern is illustrated in Fig. 1. A few or, perhaps, even one "worst defect cluster" can limit the critical current in a long tape [10]-[12]. The worst defect cluster is defined as a line of defects (strongly misaligned grains perhaps) that maximally blocks the flow of transport current. If $p$ is the probability of a grain misalignment, the normalized probability of finding a defect cluster of size $n$ is [12]

$$P(n) = (1-p)p^n . \qquad (5)$$

The number of such clusters in a tape of length L and width W is given by

$$M(n) = \frac{LW}{d^2}(1-p)p^n , \qquad (6)$$

where $d$ is the average grain size. The largest cluster is determined by the condition $M(n_{max}) = 1$, so that

$$n_{max} \approx \ln(LW(1-p)/d^2)/\ln(1/p) \qquad (7)$$

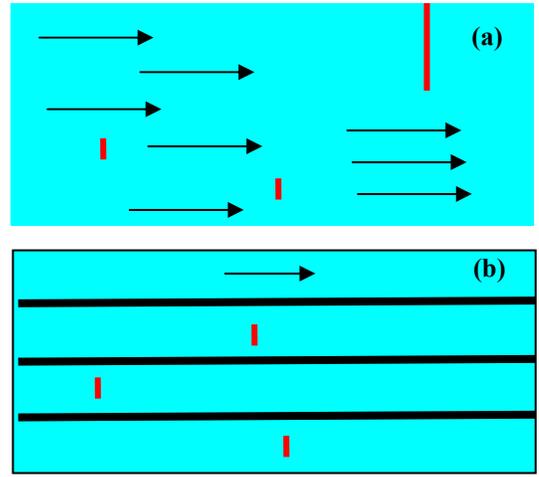

Fig. 1. (a) Sketch of a uniform superconducting tape with the "worst-defect-clusters" (vertical lines) blocking the flow of transport current. (b) Sketch of a multifilament tape with worst-defect-clusters. The horizontal lines indicate non-superconducting barriers between the stripes.

Within the critical current model the critical transport current is restricted by the cross section of the film not blocked by the largest worst defect cluster [11]:

$$I_c(L) \approx J_c(W - dn_{max}) \equiv I_c(0) - I'\ln(LW/d_0^2)) \qquad (8)$$

Here $I_c(L)$ is an "end-to-end" critical current and its logarithmic decay with length is determined by the increasing probability of appearance of greater and greater worst defect cluster. The decay rate is $I' = -\partial I_c / \partial \ln(L) = J_c d / \ln(1/p)$. If we measure the critical current on a small scale $L_0$ along the tape of length L, most measured values of $I_c(L_0)$ will be greater than $I_c(L)$, and only a few or just one small section will contain the largest defect which will determine the end-to-end critical current given by Eq. (8). One can rewrite (8) as

$$I_c(L) = I_c(L_0) - I'\ln(L/L_0) \qquad (9)$$

and define the length

$$L_{1/2} = L_0 \exp\left\{\frac{I_c(L_0)}{2I'}\right\} \qquad (10)$$

as the length of the tape for which the end-to-end critical current decays to half of its small scale value.

A summary of the world-wide results in the development of coated superconductors compiled by Los Alamos National Laboratory [13] indicates a roughly logarithmic decay of the critical current with length with $L_{1/2} \cong 1-10$ m if $L_0 \cong 1-10$ cm. The approximately logarithmic decay of $I_c(L)$ was obtained earlier by Rutter and Goyal from a statistical model of grains misalignment [14].

Now we can apply these simple estimates to the N-filament tape shown schematically in Fig. 1(b). The largest cluster in a given stripe is determined by the condition

$$n_{max} \approx \ln(LW_n(1-p)/d^2)/\ln(1/p), \qquad (11)$$

so that the critical current in a single stripe

$$I_{c,n}(L) \approx J_c(W_n - dn_{max}) \equiv I_{c,n}(0) - I'\ln(LW_n/d_0^2)) \qquad (12)$$

with the same rate of decay as in (8). Therefore, the total critical current in the multifilament tape is

$$I_c^{st}(L) = NI_{c,n}(L) = I_c(0) - NI'\ln(LW/Nd_0^2)) \quad (13)$$

Comparing this with (8), one can immediately see that the deterioration of the current-carrying capacity with length in a multifilament tape is tremendously accelerated. This is due to the cumulative effect of smaller, and therefore much more frequent, defects appearing in each stripe than the one determined by condition (7), see Fig. 1(b). Similar to Eqs. (9) and (10) we can introduce

$$I_c^{st}(L) = I_c^{st}(L_0) - NI'\ln(L/L_0); \quad L_{1/2}^{st} = L_0 \exp\left\{\frac{I_c^{st}(L_0)}{2NI'}\right\}. \quad (14)$$

As an example, let us take that in a uniform tape the "half-length" given by Eq. (10), $L_{1/2} = 1$ km when $L_0 = 10$ cm. This goal has not been achieved yet in practice, but it might be achievable. In the 20-filament tape with the same rate of decay the half-length

$$L_{1/2}^{st}(N) = L_0(L_{1/2}/L_0)^{\gamma/N} < 16 \text{ cm}, \quad (15)$$

where $\gamma = I_c^{st}(L_0)/I_c(L_0) < 1$.

A more refined treatment of the influence of defects on the critical current along the lines of [15], as well as a distinct possibility that the probability of appearance of defect clusters may not be exponential, as in (5), but algebraic might improve a pessimistic outlook on the prospects of manufacturing the long lengths of multifilament conductors that follows from (13)-(15).

On a purely empirical grounds, one can suggest that the Weibull distribution

$$P_f = 1 - \exp\{-\kappa LN^\nu I^m\}, \quad (16)$$

widely used in the statistical analysis of material failure, can also be applied to this problem. Here $P_f$ is the probability of conductor failure (reversible or permanent) and exponents $\nu$, $m$ and parameter $\kappa$ are empirical constants. Eq. (16) is based on the observation that in large systems with several risk factors for failure, the increase in one of them can be mitigated by decreasing the others and, therefore, the probability of failure has to be a function of an algebraic combination of the risk factors, rather than each of them individually. This statement is not universal. For example, if the current exceeds the critical value, appropriately defined for a given system, the failure – quench or burn-out – cannot be prevented by reduction of length or the number of filaments. The scaling described by Eq. (16) is applicable only when the probability of failure is small and all risk factors are well below their critical values – by definition the normal regime of operation.

The functional form of the probability is not important because it only needs to describe a step-like behavior with a threshold beyond which the chance of failure drastically increases. Since the goals of the loss reduction require a very large number density of stripes [7], $N \cong 10^2 - 10^3$ stripes/cm, the safe range of the length–current envelope $LI^m \propto 1/N^\nu$ may become unacceptably narrow in multifilament conductors.

## IV. MULTIPLY CONNECTED TAPES.

Fig. 2 shows two possible patterns for the superconducting layer that may alleviate the previously described limitations, while retaining an ability to reduce the magnetization losses.

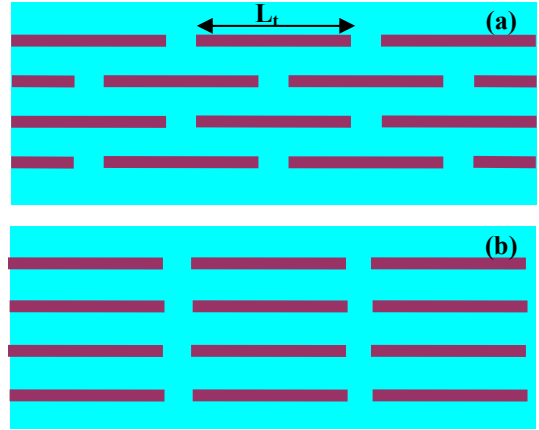

Fig. 2. (a) Sketch of two multiply connected patterns: (a) alternating bridges (brickwall pattern). The current transfer length $L_t$ is the distance between successive bridges. (b) Continuous bridges across the total width of the tape (fishnet pattern). The horizontal lines indicate non-superconducting barriers between the stripes.

The minimum width of the superconducting bridges between the stripes should be at least half the width of an individual stripe to allow the transport current from a blocked stripe to be diverted into the rest of the tape without raising the level of current criticality in the bridge above the bulk average.

Now a small defect can block a given stripe only over the distance between the successive bridges – the current transfer length. Therefore, instead of (13) the end-to-end critical current

$$I_c^{mc}(L) \approx I_c^{st}(L_t) - I'\ln(L/L_t), \quad (17)$$

where $I_c^{mc}(L)$ is the end-to-end critical current of multiply connected tape of length $L \gg L_t$ and

$$I_c^{st}(L_t) = I_c(0) - NI'\ln(L_t W/Nd_0^2).$$

As long as the critical current in a multifilament section of length $L_t$, $I_c^{st}(L_t)$, can be kept sufficiently high, the overall decay rate remains the same as in (9), and the useful length of such tape can be comparable to that of the uniform conductor. The Weibull failure distribution can be modified as follows:

$$P_f(L,I) = 1 - \exp\{-\kappa L_t N^\nu I^m\}\exp\{-\kappa(L-L_t)I^m\}. \quad (18)$$

*Additional Magnetization Losses*

The bridges that provide connectivity between the filaments also increase the amount of power loss due to supercurrent induced by the alternating magnetic field. The multifilament tape needs to be twisted in order to limit the coupling losses, (3). Fig. 3 illustrates the relation between a long twisted tape and experimental samples used for measuring the magnetization losses. The length of the samples L imitates a half twist pitch.

The losses in the superconducting bridges are determined by the electric field directed perpendicular to the stripes [3], $E_\perp \cong Bfx$, where x is the distance from the centerline indicated by the arrow in Fig. 3(b). If $\Delta$ is the width of the bridge, the

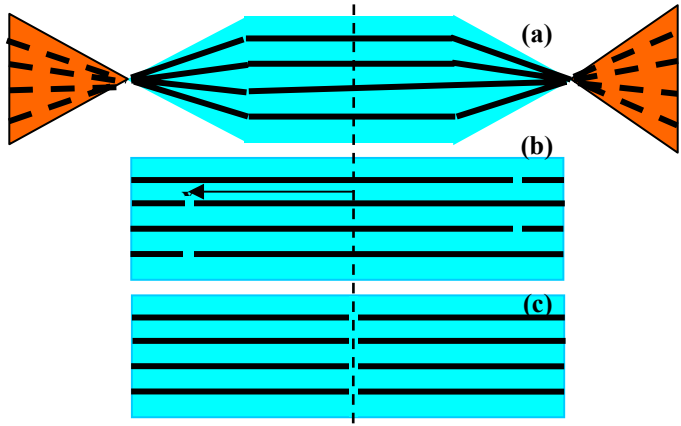

Fig. 3. (a) Sketch of twisted multifilament tape: shown is the area of the tape normal to the applied magnetic field – front side in the center of the sketch and back side at the ends. (b) and (c) are samples that imitate half of the twist pitch. (b) part of the brickwall pattern. The arrow indicates position of the bridge relative to the centerline. (c) Continuous bridge (part of the fishnet pattern) along the centerline.

power loss per unit length in the sample shown in Fig. 3(b) is given by

$$Q_s^{br} \approx J_c |E_\perp| \frac{\Delta}{L} W = I_c Bf \frac{x\Delta}{L}. \quad (19)$$

If we take $\Delta = W_n/2$, the net loss due to supercurrent induced in the stripes, (2), and bridges, (19), is

$$Q_s^{mc} = Q_h^{st} + Q_s^{br} = I_c Bf W_n + I_c Bf \frac{x\Delta}{L} = I_c Bf W_n \left(1 + \frac{x}{2L}\right) \quad (20)$$

Since $x < L/2$, the bridges shown in Fig. 3(b) increase the hysteresis loss by less than 25% in comparison with the sample where stripes are completely disconnected. It should be noted that the sample shown in Fig. 3(b) is equivalent to a twisted tape in Fig. 2(a) with the current transfer length $L_t = 4x$. For example, a 10 cm long sample with the alternating bridges located at distance $|x| = 4$ cm from the centerline is equivalent to a twisted tape with the twist pitch 20 cm and current transfer length 16 cm.

The most favorable location of the bridge is along the centerline as shown in Fig. 3(c). Since the magnetic field penetrates freely through the slits, the amount of heat generated in the bridge can be estimated using (1), and treating the bridge as a stripe whose *length* is $W$ and width is $\Delta$. Thus, the amount of heat released in the bridge per unit length of the sample

$$Q_{br} \approx \frac{J_c \Delta^2 W Bf}{L} \equiv I_c W_n Bf \frac{\Delta^2}{W_n L}. \quad (21)$$

If we take $\Delta = W_n/2$, the contribution of the centrally located bridge to losses is negligible in comparison with that of the stripes, Eq. (2). Moreover, the width of the bridge can be increased to be equal or even greater than $W_n$ without paying a significant penalty in terms of increased losses. This will make such a bridge a more robust tool for mixing and redistributing the transport current between the stripes.

The bridge location shown in Fig. 3(c) corresponds to the current transfer length $L_t = L$ – half of the twist pitch. The brickwall pattern, Fig. 2(a) with $L_t = 2L$ can also be arranged so that the bridges will be lined up along the centerline. However, (20) suggests that even if such a transfer length will prove to be too long to be effective, there is room for compromise and one can place several bridges per half of the twist pitch before the hysteresis loss increases by 100% in comparison with the disconnected multifilament tape.

## V. SUMMARY

We have presented arguments that a multifilament coated superconductor in which the superconducting stripes are not connected with each other may prove to be difficult to produce in substantial length because of the cumulative effect of relatively small defects and defects clusters on its current-carrying capacity. The prospects can be improved if we allow a sparse network of superconducting bridges to connect the stripes as a means to divert the supercurrent from the damaged areas. The amount of additional losses incurred in the bridges is tolerable as long as the distance between them is comparable to the length of the twist pitch.

ACKNOWLEDGMENT

We would like to thank C. E. Oberly, M. Sumption, N. Amemiya, T. Haugan, S. Sathiraju, and C. Varanasi for useful conversations and discussions.